\documentclass[aps,prb,twocolumn,superscriptaddress,floatfix]{revtex4-2}
\usepackage{graphicx,graphics}
\usepackage{dcolumn}
\usepackage{amsmath,amssymb,amsfonts,nccmath}
\usepackage{latexsym,verbatim}
\usepackage{bm}
\usepackage{color}
\usepackage{ulem}
\usepackage{soul}
\usepackage{braket}
\usepackage[percent]{overpic}
\usepackage[breaklinks=true,colorlinks,citecolor=blue,linkcolor=blue,urlcolor=blue]{hyperref}
\usepackage{lipsum}
\usepackage{enumitem}
\usepackage[T1]{fontenc}

\usepackage{orcidlink}

\usepackage{xprintlen}

\makeatletter
\renewcommand\@make@capt@title[2]{%
 \@ifx@empty\float@link{\@firstofone}{\expandafter\href\expandafter{\float@link}}%
  {\textbf{#1}}\@caption@fignum@sep#2\quad
}
\makeatother

\graphicspath {{Images/}}

\begin{document}
\title{Current phase relation in a planar graphene Josephson junction with spin-orbit coupling}
%

%
\author{Federico Bonasera~\orcidlink{0000-0002-2095-6646}}
\affiliation{Dipartimento di Fisica e Astronomia "Ettore Majorana",  Universit\'a di Catania, Via S. Sofia, 64, I-95123 Catania,~Italy.}
\affiliation{INFN, Sez. Catania, I-95123 Catania,~Italy.}
\author{Giuseppe A. Falci~\orcidlink{0000-0001-5842-2677}}
\affiliation{Dipartimento di Fisica e Astronomia "Ettore Majorana",  Universit\'a di Catania, Via S. Sofia, 64, I-95123 Catania,~Italy.}
\affiliation{INFN, Sez. Catania, I-95123 Catania,~Italy.}
%
\author{Elisabetta Paladino~\orcidlink{0000-0002-9929-3768}}
\affiliation{Dipartimento di Fisica e Astronomia "Ettore Majorana",  Universit\'a di Catania, Via S. Sofia, 64, I-95123 Catania,~Italy.}
\affiliation{INFN, Sez. Catania, I-95123 Catania,~Italy.}

%
\author{Francesco M.D. Pellegrino~\orcidlink{0000-0001-5425-1292}}
\affiliation{Dipartimento di Fisica e Astronomia "Ettore Majorana",  Universit\'a di Catania, Via S. Sofia, 64, I-95123 Catania,~Italy.}
\affiliation{INFN, Sez. Catania, I-95123 Catania,~Italy.}
%
\begin{abstract}
We study a graphene Josephson junction where the inner graphene layer is subjected to spin-orbit coupling by proximity effect. This could be achieved, for example, by growing the graphene layer on top of a transition metal dichalcogenide, such as WS$_2$. Here, we focus on the ballistic, wide, and short junction limits and study the effects of the spin-orbit interaction on the supercurrent. In particular, we analyze the current phase relation using an analytical approach based on the continuum model. We find combinations of types of spin-orbit coupling that significantly suppress the supercurrent by opening a gap in the graphene band structure. At the same time, other combinations enhance it, acting as an effective spin-valley resolved chemical potential. Moreover, we find that a strong Rashba spin-orbit coupling leads to a junction with a highly voltage tunable harmonic content.
\end{abstract}
%
\maketitle

\section{Introduction}
\label{sec:intro}

Superconducting-based systems are currently among the most promising platforms for quantum information processing \cite{deLeon_2021_a}. Specifically, superconducting qubits have been extensively researched and utilized as components in quantum computing \cite{Kjaergaard_2020_a,Huang_2020_a,Ezratty_2023_a,Falci_2024_a,Balandin_2024_a,Falci_1991_a}. 
They encode information in electrical circuits whose main element is the Josephson junction (JJ).
%
Recently, the semiconductor-based JJ has opened up new possibilities for superconducting qubits with a gate-tunable ratio between the Josephson and the charging energies, $E_{\rm J}/E_{\rm C}$. Implementations of this kind have been made with a variety of materials \cite{Larsen_2015_a,Kringoj_2018_a,Lee_2019_a,Antony_2021_a,Sagi_2024_a,Kroll_2018_a}. Semiconductor-based JJs often work in the quasi-ballistic regime and are characterized by few high transmission channels that produce a skewed current phase relation (CPR) with higher harmonic content \cite{Beenakker_1991_a,Spanton_2017_a,Kringoj_2018_a}.
%
%
There is growing literature on these kinds of JJs. For example, high-transmission superconducting JJs are at the base of multiterminal topological effects \cite{vanHeck_2014_a,Riwar_2016_a,Pientka_2017_a};
%
%
they were also proven to suppress charge dispersion in transmon-like qubits \cite{Kringoj_2020_a};
%
%
moreover, in a SQUID configuration, it was demonstrated that the higher harmonics could be used to implement high-efficiency superconducting diodes \cite{Souto_2022_a,Valentini_2024_a} and also parity protected $\sin (2\phi)$ qubits \cite{Larsen_2020_a,Gyenis_2021_a,Valentini_2024_a}.

Graphene is one of the most promising materials for such applications. In fact, thanks to advances in the fabrication of encapsulated graphene, it is now possible to produce high-quality superconducting graphene heterostructures \cite{Dean_2010_a,Mayorov_2011_a,Wang_2013_a}. These systems show ballistic transport, gate-tunable supercurrents, and forward-skewed CPRs due to high transmission channels \cite{Calado_2015_a,BenShalom_2016_a,Borzenets_2016_a,English_2016_a,Nanda_2017_a,Allen_2017_a,Pellegrino_2020_a}.
%
%
One of the most recent observations of a $\sin(2\phi)$ CPR was made with a graphene-based SQUID \cite{Messelot_2024_a}.
Recently, it was found that when encapsulated with transition-metal dichalcogenides, graphene can acquire a strong spin-orbit coupling (SOC) by proximity effect \cite{Avsar_2014_a,Mendes_2015_a,Gmitra_2015_a,Wang_2015_b,Wakamura_2018_a,Sun_2023_a}.
%
%
In general, the interplay of SOC and superconductivity can lead to exotic phenomena such as the anomalous Josephson effect \cite{Yokoyama_2014_a,Szombati_2016_a,Mayer_2020_a},
%
%
%
the triplet \cite{Reeg_2015_a,Beiranvand_2016_a,Beiranvand_2017_a},
%
%
%
or the finite-momentum \cite{Hart_2017_a,Guo_2022_a} pairing superconductivity, and the sought-after topological superconductivity \cite{Fu_2008_a,Sau_2010_a}. In graphene, the presence of a strong SOC was also linked to the extreme robustness of the supercurrent against high magnetic fields \cite{Wakamura_2020_a}.

Motivated by the previous observations, we study the CPR through a ballistic graphene JJ (GJJ), in which the inner graphene layer is subjected to the SOC interaction by proximity effect. We focus on the short and wide junction limits, in which the junction length is much smaller than the coherence length, $\xi$, of the superconductors. We employ an analytical approach and use a low-energy description of graphene based on the continuum Hamiltonian~\cite{Kochan_2017_a}. Using the transfer matrix formalism, we compute the transmission probabilities through the junction and directly link them to the CPR \cite{Titov_2007_a,Pellegrino_2011_a,Beenakker_1992_a,Titov_2006_a,Beenakker_2008_a,Hagymasi_2010_a}.
%
%
The results focus on the critical current and skewness of the CPR as a function of the chemical potential, showing the effects of Kane-Mele (KM), valley-Zeeman (VZ), and Rashba SOCs, and the onsite scalar potential. We find that, depending on the modifications of the band structure, these different terms result in a wide range of effects on the CPR of the junction. Some of them, such as the on-site potential and the KM SOC, drastically reduce both the critical current and the skewness. Others, such as the VZ SOC act as a spin-valley-polarized chemical potential, increasing both the critical current and the skewness. Interestingly, the Rashba SOC, while having little effect on the critical current of the CPR, produces heavy swings in its skewness, allowing for a tunable harmonic content.

This paper is organized as follows. In Sec.~\ref{sec:model} we introduce the continuum Hamiltonian of the inner graphene region. We also introduce the transfer matrix formalism, which we use to compute the transmissions through the junction, together with the formulae that link them to the CPR of the system. In Sec.~\ref{sec:results} we present the results of the study. First, we show the cases without Rashba SOC, which we were able to fully compute analytically; then, when also introducing the Rashba SOC, we resort to solving the problem numerically and show the results for some of the most experimentally relevant parameter values. Finally, a summary of the work and some final comments are included in Sec.~\ref {sec:conclusions}.

\section{Model}\label{sec:model}

\begin{figure}[t!]
\centering
\begin{overpic}[width=0.98\columnwidth,trim={0 0cm 0 0cm}]{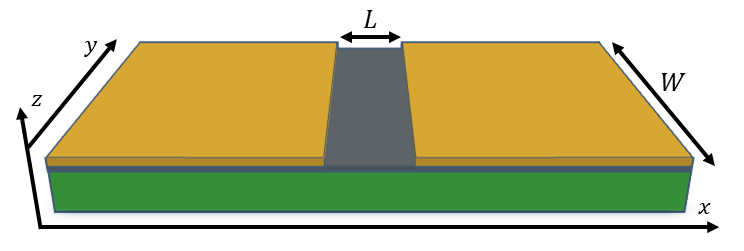}\end{overpic}
\caption{Schematic of the system. A GJJ consists of a graphene layer (gray) placed on top of a substrate (green) with superconducting leads (yellow) covering the $\left|x\right| > L/2$ regions.
}
\label{fig:setup}
\end{figure}
Figure~\ref{fig:setup} illustrates a diagram of the system: a gray graphene layer is positioned on top of a green substrate, with yellow superconducting leads that extend over the areas where $\left|x\right| > L/2$. The width of the junction in the $y$ direction is treated as infinite, $W \to \infty$.

To describe the system, we use a step-like profile for the different regions of the junction. The Hamiltonian of the inner region of the junction includes the terms induced by the substrate on the graphene layer and, when expanded to first order in the momentum $\bm{k}$ around the Dirac points, is expressed as \cite{Min_2006_a,Gmitra_2009_a,Gmitra_2015_a,Kochan_2017_a,Frank_2018_a}
\begin{equation}\label{eq:Hamiltonian}
\begin{aligned}
\mathcal{H} & = - i\hbar v \left[ \tau_z \left( \partial_x \sigma_x \right) + \partial_y \sigma_y \right] - \mu_0 + U_z \sigma_z \\ & + s_z \tau_z \left( \lambda_{\rm KM} \sigma_z + \lambda_{\rm VZ} \right) - \lambda_{\rm R} \left( s_y \tau_z \sigma_x + s_x \sigma_y \right),
\end{aligned}
\end{equation}
which acts on an $8$-dimensional spinorial space in which $\bm{\sigma}$, $\bm{\tau}$ and $\bm{s}$ are the three Pauli matrices that describe the sublattice ($A,B$), valley ($\bm{K},-\bm{K}$) and spin $z$ projection ($\uparrow,\downarrow$) degrees of freedom, respectively. In Eq.~\eqref{eq:Hamiltonian}, $\mu_0$ is the Fermi level of the graphene layer, $U_z$ represents the intensity of the staggered onsite scalar potential on the $A,B$ sublattices, $\lambda_{\rm KM}$ denotes the intensity of the KM SOC \cite{Kane_2005_a,Kane_2005_b}, $\lambda_{\rm VZ}$ refers to the strength of the valley Zeeman (VZ) SOC, and $\lambda_{\rm R}$ represents the magnitude of the Rashba SOC. All of the aforementioned terms come from a reduction in symmetry due to the presence of the substrate \cite{Kochan_2017_a}. In particular, the onsite scalar potential comes from the breaking of the $A,B$ sublattice symmetry. The other SOC terms come, instead, from the breaking of the $z$ inversion symmetry.

We focus on the short junction limit, $L \ll \xi \sim \hbar v/\Delta_0$, where $\xi$ is the coherence length of the superconductors and $\Delta_0$ is the superconducting gap parameter. In this regime, the supercurrent is essentially carried by  Andreev Bound States (ABSs), which are intra-gap states, $\left| E_p \right| < \Delta_0$, and are localized inside the inner region of the junction~\cite{Beenakker_1992_a,Titov_2006_a,Beenakker_2008_a}. The energy spectrum of the ABSs can be linked to the transmission probability, $\mathcal{T}_p$, of normal electrons passing through a ballistic graphene stripe via the following formula \cite{Beenakker_1992_a,Titov_2006_a,Beenakker_2008_a}
\begin{equation}\label{eq:ABS_Spectrum}
E_p = \Delta_0 \sqrt{ 1 - \mathcal{T}_p \sin^2 \left(\phi/2\right)  }~,
\end{equation}
where $\phi$ is the superconducting phase difference between the superconductors, and $p$ is a set of indices that characterize an ABS. The supercurrent carried by these states, at zero temperature, can be computed as \cite{Beenakker_1992_b,Beenakker_1992_a,Titov_2006_a,Beenakker_2008_a,Hagymasi_2010_a}
\begin{equation}\label{eq:Full_Supercurrent_Beenakker}
I\left(\phi\right) = - \frac{2e}{\hbar} \sum_{E_p>0} \frac{dE_p}{d \phi}.
\end{equation}
%

Based on Eq.~\eqref{eq:ABS_Spectrum}, we calculate the normal-state transmission amplitude, $\mathcal{T}_p$, as a function of the SOC terms. 
For this aim, we start from the setup in Fig.~\ref{fig:setup} by considering lateral leads in the normal metallic limit and follow the same procedure as in Refs.~\cite{Titov_2007_a,Pellegrino_2011_a}. We solve the stationary Schr\"{o}dinger equation in all three regions and ensure that the wave function is continuous throughout the device~\cite{Pellegrino_2022_a,Vacante_2025_a}. Specifically, we apply the transfer matrix method within the central region to enforce wave function matching.
%
%
Using periodic boundary conditions along the $y$-direction, we write the generic eigenfunction, labeled by the wavevector $k$ and the energy $E$, in the central region as
\begin{gather}\label{eqw:TrasnferMatrix_Def}
\psi_{k,E}\left(x,y\right) = e^{iky} T \left(k,E;x\right) \tilde{\psi}_{k, E}\left(-L/2\right)~,
\end{gather}
where the evolution matrix $T$ satisfies the equation below
\begin{widetext}
\begin{equation}\label{eq:TransferMatrix_SE}
\begin{aligned}
\frac{d T\left( k, E ; x \right)}{dx} & = \frac{\mathcal{A}}{\hbar v} T \left( k, E ; x \right),\\
{\cal A} & \equiv i [ \left(\mu_0 + E \right) \tau_z - \lambda_{\rm VZ} s_z ] \sigma_x - [\lambda_{\rm KM} { s_z} + U_z \tau_z] \sigma_y + \lambda_{\rm R} \left( i s_y - s_x \tau_z \sigma_z \right) +  \hbar v k \tau_z \sigma_z,
\end{aligned}
\end{equation}
\end{widetext}
which, together with the condition $T \left( k, E, -L/2 \right) = 1$, has the formal solution
\begin{equation}\label{eq:TransferMatrix_Formal}
T\left( k, E; x \right) = \exp{\left[ \frac{\mathcal{A}}{\hbar v} \left( x + L/2 \right) \right]}.
\end{equation}
%
%
In a GJJ,  the typical energy scale for the ABSs is $E \sim \Delta_0$, and in the short junction regime $\Delta_0 \ll L/(\hbar v)$, so we can focus on $\mathbb{T} \left(k;x\right)=T(k,0;x)$.
%
In the presence of metallic leads, the evolution matrix is related to the transfer matrix by \cite{Titov_2007_a,Pellegrino_2011_a}
\begin{equation}
M(k)=\mathcal{Q}^{-1} {\mathbb{T}}(k,L/2) \mathcal{Q},
\end{equation}
with $\mathcal{Q} = \frac{1}{\sqrt{2}} \left( \sigma_x + \sigma_z \right) $.
Moreover, the elements of the transfer matrix are linked to those of the scattering matrix across the region as
\begin{equation}\label{eq:TransferMatrix_To_ScatteringMatrix}
M(k)=
\begin{pmatrix}
\left(t^\dagger_{12} \left(k\right)\right)^{-1} & r_{22}\left(k\right) \left( t_{21} \left(k\right)\right)^{-1} \\
- r_{11} \left(k\right) \left( t_{21}\left(k\right) \right)^{-1} & \left( t_{21} \left(k\right)\right)^{-1},
\end{pmatrix} ,
\end{equation}
%
with $t_{ij}\left(k\right)$ ($r_{ii}\left(k\right)$) the transmission (reflection) amplitude matrix from lead $i$ to lead $j$ ($i$). 
The transmission probabilities are then found as the eigenvalues of the hermitian matrix $ t_{12}^\dagger \left(k\right)t_{12}\left(k\right)$.

\begin{figure*}
\centering
\begin{overpic}[width=0.99\columnwidth,trim={0 0cm 0 0cm}]{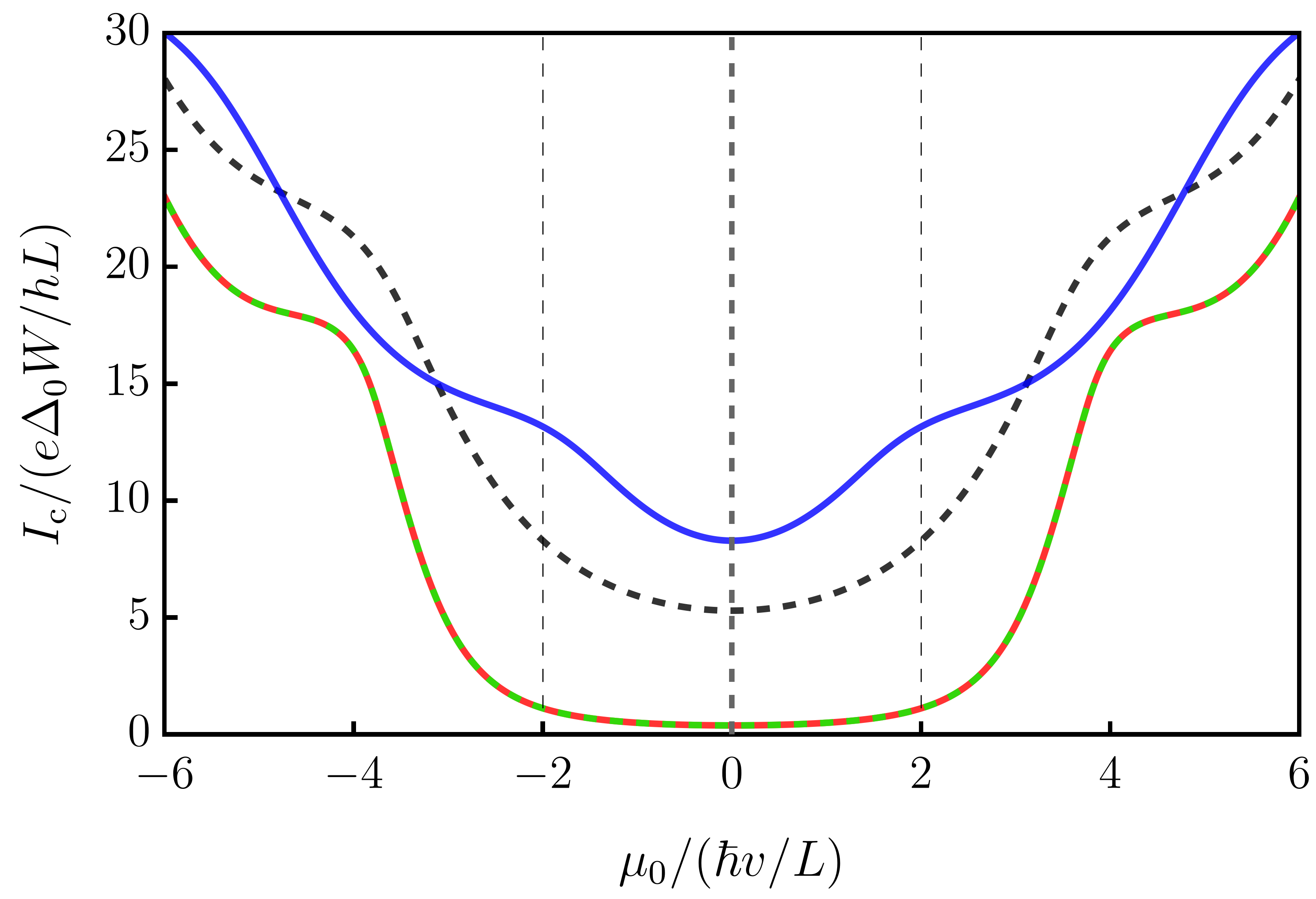}\put(0,70){a)}\end{overpic}
\begin{overpic}[width=0.99\columnwidth]{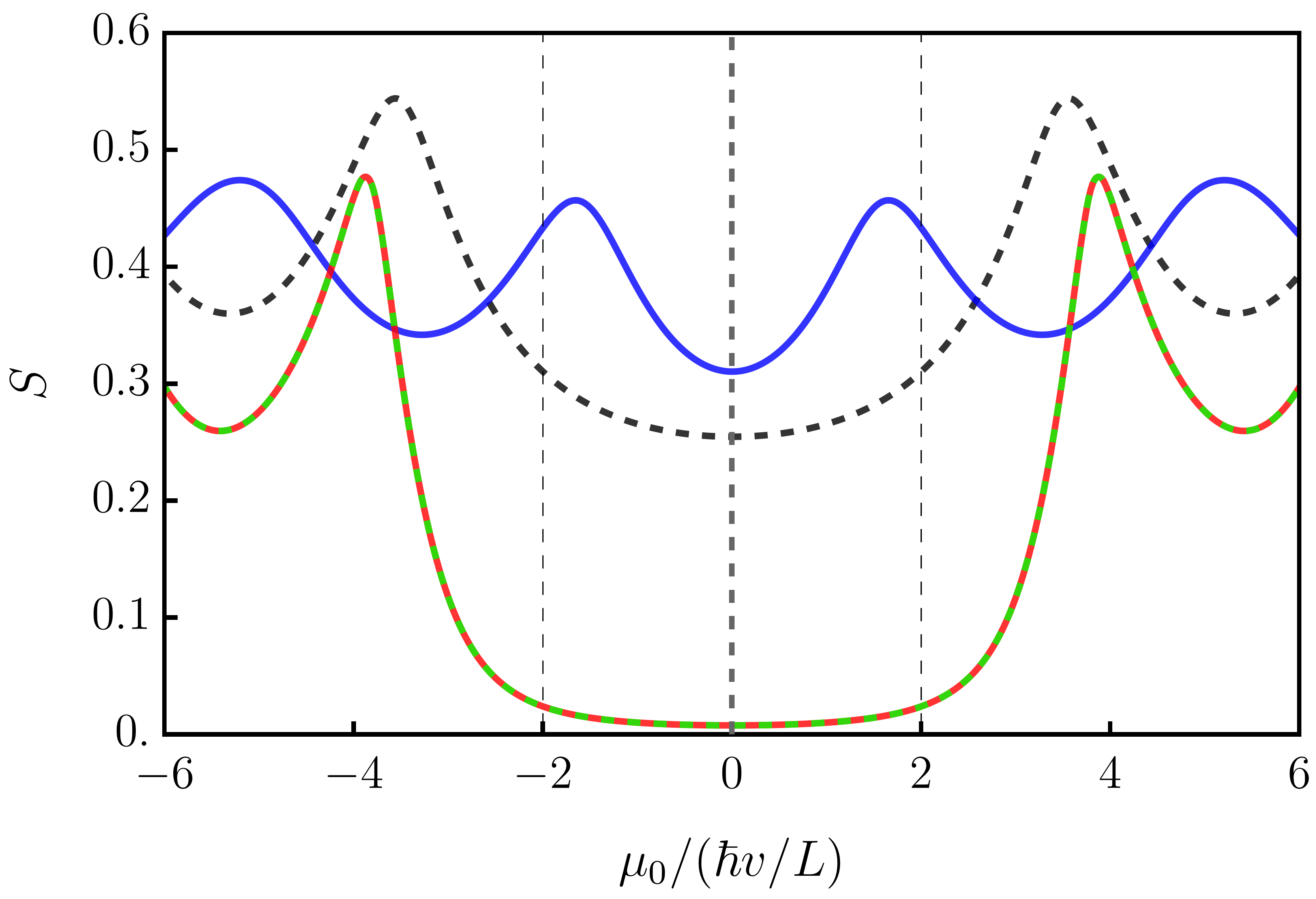}\put(0,70){b)}\end{overpic}
\\[2em]
\begin{overpic}[width=0.99\columnwidth,trim={0 0cm 0 0cm}]{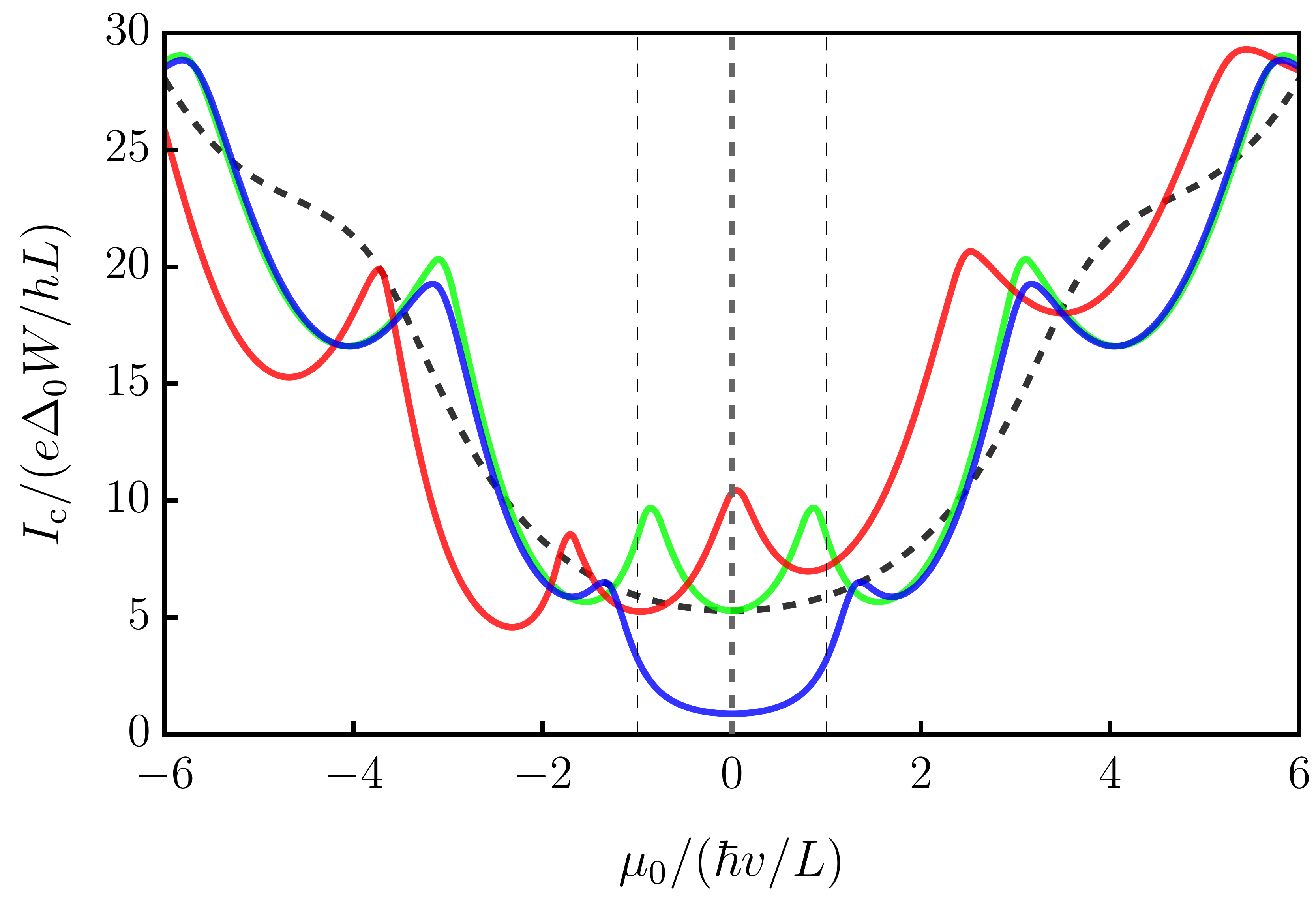}\put(0,70){c)}\end{overpic}
\begin{overpic}[width=0.99\columnwidth]{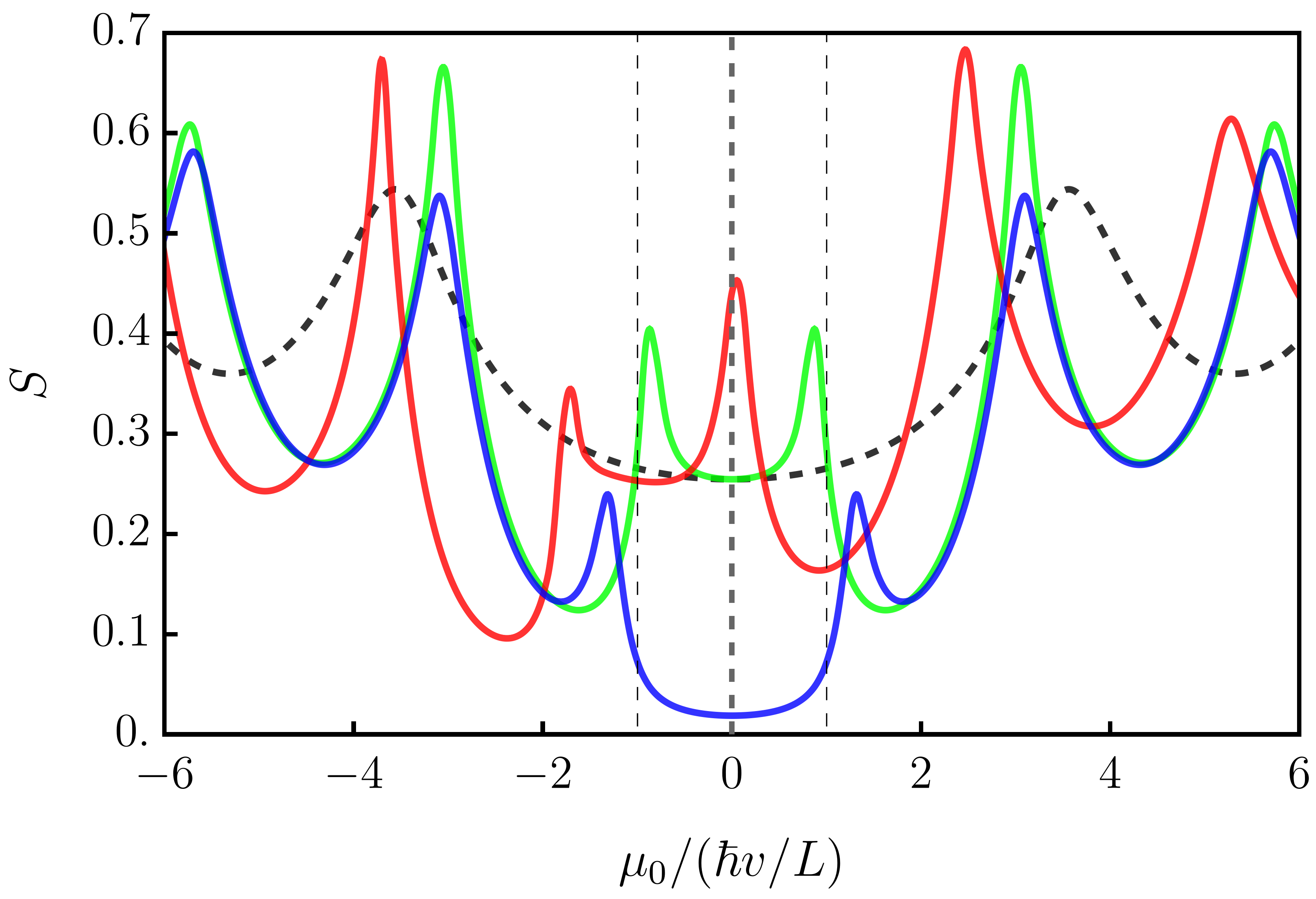}\put(0,70){d)}\end{overpic}
\caption{a) critical current $I_{\rm c}$ and b) skewness $S$, as a function of the chemical potential $\mu_0$ for each substrate-induced term independently, when $\lambda_{\rm R} = 0$.
For reference, dashed black refers to the intrinsic case (substrate-independent terms). Solid lines consider the effect of the VZ SOC $\lambda_{\rm VZ} = 2 \hbar v/L$ (blue), and of the KM SOC $\lambda_{\rm KM} = 2 \hbar v/L$ (red). The dashed green line denotes the case with the staggered onsite potential $U_z = 2 \hbar v/L$.
c) critical current $I_{\rm c}$ and d) skewness $S$, as a function of the chemical potential $\mu_0$ for different parameter sets given a non-zero $\lambda_{\rm R}$. Again, the dashed black line denotes the intrinsic case for reference, the solid green line denotes the effect of a high Rashba SOC, $\lambda_{\rm R} = 5\hbar v/L$, the solid blue one the case with a strong Rashba and a smaller VZ SOC, $\lambda_{\rm R} =5\hbar v/L, \lambda_{\rm VZ} = \hbar v/L$, and the solid red one that with a strong Rashba and a smaller KM SOC, $\lambda_{\rm R} =5\hbar v/L, \lambda_{\rm KM} = \hbar v/L$.
All the energy parameters are scaled in units of $\hbar v/L$ and the critical current is in units of $e\Delta_0 W /(h L)$.}
\label{fig:NoRashba_1}
\end{figure*}

\section{Results}
\label{sec:results}

In this Section we show the results for the supercurrent flowing through the GJJ. 

Firstly, we neglect the Rashba SOC, setting $\lambda_{\rm R} = 0$. 
This simplification allows us to handle the transfer matrix analytically, thereby clarifying the influence of the KM and VZ SOC terms on the supercurrent. In the last part of this Section, we incorporate the Rashba SOC and tackle the problem using numerical methods~\cite{Mathematica_141}.


In the absence of $\lambda_{\rm R}$, the spin $z$ projection and the valley index are good quantum numbers. So, Eq.~\eqref{eq:TransferMatrix_SE} can be expressed for the $4$-dimensional spin-valley subspaces as
\begin{equation}\label{eq:TransferMatrix_SE_Pauli}
\frac{d \mathbb{T}_{s \tau} \left(k; x \right)}{d x} = { \tau } \left[i \alpha_{s \tau} \sigma_x - \gamma_{s \tau} \sigma_y + k \sigma_z \right] \mathbb{T}_{s \tau} \left( k; x \right),
\end{equation}
where the $s$ ($\tau$) index takes on the values $s\in\{+,-\}$ ($\tau\in\{+,-\}$), corresponding to the spin $z$-component (valley) $\uparrow$ and $\downarrow$ ($\bm K$ and $\bm K'$), respectively.
We also defined the following renormalized parameters 
\begin{subequations}\label{eq:TransferMatrix_SE_Pauli_Explicit_Terms}
\begin{align}
\alpha_{s \tau} & = \frac{ \mu_0 - s \tau \lambda_{\rm VZ} }{ \hbar v}, \label{eq:TransferMatrix_SE_Pauli_Explicit_Terms_a} \\
\gamma_{s \tau} & = \frac{ U_Z + { s \tau} \lambda_{\rm KM} }{ \hbar v}, \label{eq:TransferMatrix_SE_Pauli_Explicit_Terms_b}
\end{align}
\end{subequations}
which depend on the $s$ and $\tau$ indices by their product $s \tau$.
For each couple of $s$ and $\tau$,
we solve Eq.~\eqref{eq:TransferMatrix_SE_Pauli} in the following close analytical form
\begin{equation}\label{eq:TMatrix_Explicit_Pauli}
\begin{aligned}
\mathbb{T}_{s \tau} \left(k; x \right) & = \cos\left[ q_{s \tau} \left( x + L/2 \right)\right] \\ & + {\tau} \frac{\sin \left[ q_{s \tau} \left( x + L/2 \right) \right]}{q_{s \tau}} \left( i \alpha_{s \tau} \sigma_x + i \gamma_{s \tau} \sigma_y + k \sigma_z \right),
\end{aligned}
\end{equation}
with
\begin{equation}\label{eq:TMatrix_Explicit_Pauli_Q}
q_{s \tau} = \sqrt{ \alpha_{s \tau}^2 - \gamma_{s \tau}^2 - k^2 }.
\end{equation}
%
%

Using Eqs.~\eqref{eq:TransferMatrix_To_ScatteringMatrix}, we can then find the transmission probabilities as the eigenvalues of the hermitian matrix $t_{12}\left(k\right)t_{12}^\dagger\left(k\right)$, which are expressed as
\begin{equation}\label{eq:TransmissionProbabilities}
\mathcal{T}_{s \tau} \left(k \right) = \frac{  \alpha_{s \tau}^2 - \left( k^2 + \gamma_{s \tau}^2 \right)  }{\alpha_{s \tau}^2 - \left(k^2 + \gamma_{s \tau}^2 \right) \cos^2 \left[ L \sqrt{ \alpha_{s \tau}^2 - \left( k^2 + \gamma_{s \tau}^2 \right) } \right]}.
\end{equation}
%
%
Finally, using Eqs.~\eqref{eq:ABS_Spectrum} and \eqref{eq:Full_Supercurrent_Beenakker}, we obtain the supercurrent carried by the ABSs as
\begin{equation}\label{eq:Supercurrent_NoRashba}
\begin{aligned}
I \left( \phi \right) & = \frac{2e \Delta_0}{\hbar} \sum_{k,s,\tau} \frac{\mathcal{T}_{s \tau} \left(k \right) \sin \phi}{4 \sqrt{1-\mathcal{T}_{s \tau} \left(k \right) \sin^2 (\phi/2)}} \\
& = \frac{2e \Delta_0}{h} W \sum_{s,\tau = \pm} \int dk \frac{\mathcal{T}_{s\tau} \left(k \right) \sin \phi}{4 \sqrt{1-\mathcal{T}_{s\tau} \left(k \right) \sin^2 (\phi/2)}},
\end{aligned}
\end{equation}
where in the last line we have considered the wide junction limit ($W\to \infty$), namely the sum over $k$ is replaced by an integral $\sum_k \to  W/2\pi \int dk$.

%
%

The results shown in Eqs.\eqref{eq:TransmissionProbabilities} and 
\eqref{eq:Supercurrent_NoRashba}
~
are formally equivalent to those found for a short JJ based on intrinsic graphene \cite{Tworzydlo_2006_a,Pellegrino_2022_a}, provided the following replacements
\begin{subequations}\label{eq:ParameteresSubstitution_Pristine}
\begin{align}
\frac{\mu_0}{\hbar v} & \to \alpha_{s \tau}, \label{eq:ParameteresSubstitution_Pristine_a} \\
k^2 & \to k^2 + \gamma_{s \tau}^2. \label{eq:ParameteresSubstitution_Pristine_b}
\end{align}
\end{subequations}
In particular, $\lambda_{\rm VZ}$ determines the parameter $\alpha_{s \tau}$, which works as an effective Fermi level for the spin-valley interaction. Moreover, $\lambda_{\rm KM}$ and $U_z$ determine the $\gamma_{s \tau}$ parameter, which creates an effective spin-valley gap. A non-zero value of $\gamma_{s \tau}$ disrupts Klein tunneling at $k=0$ \cite{Klein_1929_a,Beenakker_2008_a} and can significantly decrease the total supercurrent passing through the junction for small values of $|\mu_0|$.
In particular, both $U_z$ and $\lambda_{\rm KM}$ independently open the same size gap in graphene and have the same effect on the transmission probability of Eq.~\eqref{eq:TransmissionProbabilities}, which depends quadratically on $\gamma_{s \tau}$. The difference between the two terms is that $\lambda_{\rm KM}$ opens an inverted (topological) gap, meaning that the conduction band has a different $A,B$ sublattice polarization at the two Dirac points; see Eq.~\eqref{eq:Hamiltonian} \cite{Kane_2005_a,Kane_2005_b,Kochan_2017_a,Frank_2018_a}. For this reason, when they are both present, their effect is enhanced in one spin-valley subspace, while it is decreased in the other one, as can be seen from the expression of $\gamma_{s\tau}$ in Eq.~\eqref{eq:TransferMatrix_SE_Pauli_Explicit_Terms_a}. The competition between these two terms is not trivial, and it was also recently used in bilayer graphene (where $U_z$ comes from the difference between the chemical potentials of the upper and lower layers) to probe its topological state \cite{Rout_2024_a}.

To characterize the equilibrium supercurrent, $I \left(\phi\right)$, flowing through the junction, we use the critical current, defined as its maximal value
\begin{equation}\label{eq:CriticalCurrent_Definition}
I_{\rm c} = \max_{\phi} I \left(\phi \right).
\end{equation}
For any given set of microscopic parameters, from the definition above, we call $\phi_{\rm max}$ the corresponding phase difference that maximizes the supercurrent, and we define the
supercurrent skewness as
\begin{equation}\label{eq:Skewness_Definition}
S\equiv \frac{2 \phi_{\rm max} - \pi}{\pi},
\end{equation}
which quantifies the deviation from a sinusoidal CPR.

To keep the junction dimensions general, in the following, the inverse length $1/L$ is used to scale the wavenumber $k$,  $\hbar v/L$ is used as the energy scale unit, while the critical current, $I_{\rm c}$, is considered in units of $(e\Delta_0 W /h L)$.
Fig.~\ref{fig:NoRashba_1} shows the results for a) the critical current $I_{\rm c}$ and b) the skewness $S$, as a function of the chemical potential $\mu_0$, when $\lambda_{\rm R} = 0$. The dashed black line represents intrinsic graphene, where all substrate-induced parameters are set to zero. The other lines refer to the case with a single non-zero substrate-induced parameter: for the solid blue line, we set $\lambda_{\rm VZ}=2 \hbar v/L$, for the solid red one $\lambda_{\rm KM}=2 \hbar v/L$, and, finally, for the dashed green one $U_z=2 \hbar v/L$. 
\begin{figure*}
\begin{minipage}{\columnwidth}
\begin{overpic}[width=0.98\columnwidth,trim={0 0cm 0 0cm}]{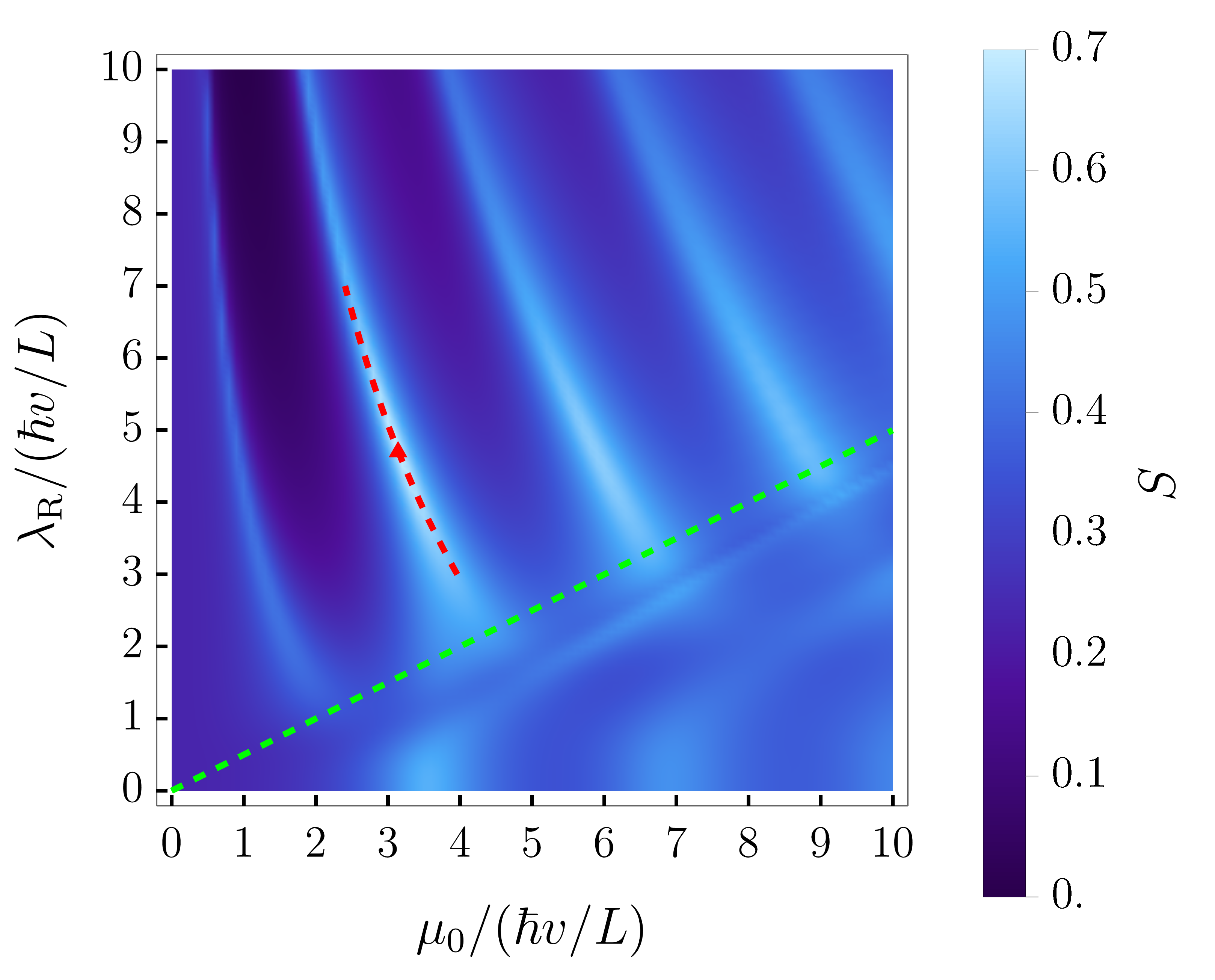}\put(0,77){a)}\end{overpic}
\end{minipage}
\begin{minipage}{\columnwidth}
\begin{overpic}[width=\columnwidth,trim={0 0cm 0 0cm}]{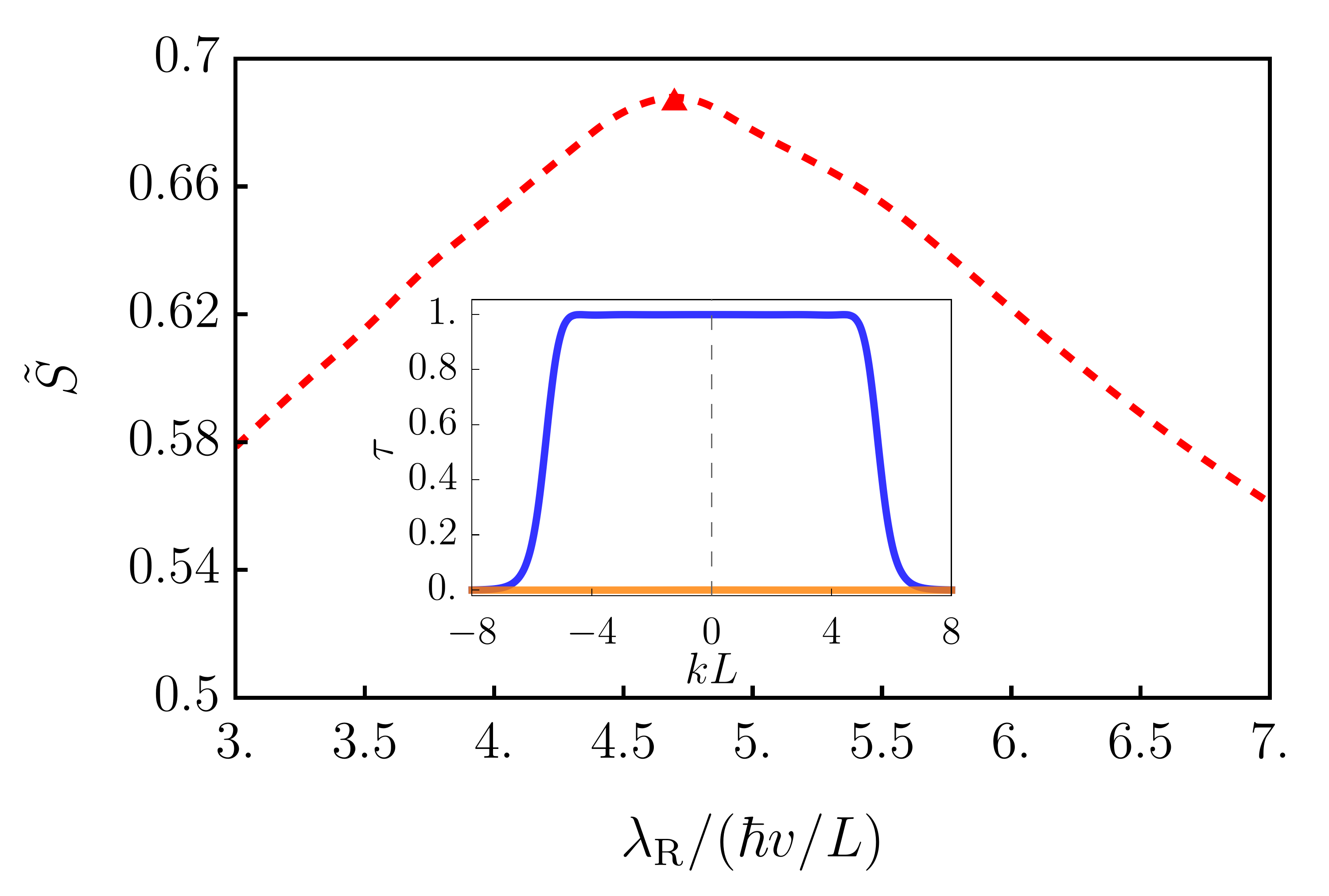}\put(0,70){b)}\end{overpic}
\end{minipage}
\caption{a) color map of $S(\lambda_{\rm R},\mu_0)$. Here, the green dotted line refers to the condition $\mu_0 = 2\lambda_{\rm R}$, where the Fermi level coincides with the bottom of the graphene upper band. The red dashed line represents the maximum of the skewness, $\tilde{S}(\lambda_{\rm R})$, within the range of the Rashba SOC $\lambda_{\rm R}\in [3,7] \hbar v/L$.
b) maximum skewness, $\tilde{S}(\lambda_{\rm R})$, as a function of $\lambda_{\rm R}$; the global maximum (red triangle) is well visible for the optimal value of $\lambda_{\rm R \blacktriangle}\approx 4.7 \hbar v/L$. The inset shows the peculiar, nearly step-like transmission probability as a function of $k$ occurring at $\lambda_{\rm R \blacktriangle}$. All energies are scaled in units of $\hbar v/L$ and the wavenumber $k$ is represented in units of $1/L$.
}
\label{fig:MaxSkewRashba}
\end{figure*}
In particular, Figs.~\ref{fig:NoRashba_1}~a) and b) clearly show that the gap opening due to the terms $U_z$ and $\lambda_{\rm K}$, in the $|\mu_0| \leq |\lambda_{\rm KM}|,|U_z|$ region indicated by the vertical dashed lines, drastically reduces both the critical current and, by eliminating the transparent modes due to Klein tunneling at $k=0$, also the skewness of the CPR.
In addition, a nonzero $\lambda_{\rm VZ}$ term enhances the critical current when the chemical potential is close to $\mu_0 \approx 0$. Also, because of the form of $\alpha_{s \tau}$ in Eq.~\eqref{eq:TransferMatrix_SE_Pauli_Explicit_Terms_a}, the critical current stays approximately constant for $|\mu_0| \lesssim |\lambda_{\rm VZ}|$.

%

In the second part of this Section, we include a nonzero Rashba SOC, $\lambda_{\rm R} \neq 0$.  Here, the spin $z$-projection is no longer a good quantum number and is involved in the dynamics. For this reason, with a finite Rashba SOC, we numerically solve Eq.~\eqref{eq:TransferMatrix_SE} for some specific sets of parameters. Recent works suggest that in a graphene monolayer proximitized by a substrate with a large SOC, the Rashba term is typically the dominant one \cite{Wang_2016_a,Wakamura_2018_a,Sun_2023_a}. 
Thus, among all the possible combinations of SOC parameters, from now on we focus on a high Rashba SOC, $\lambda_{\rm R}$, combined with a smaller secondary one of the KM kind, $\lambda_{\rm KM}$, or VZ one, $\lambda_{\rm VZ}$. 
Figure \ref{fig:NoRashba_1} shows c) the critical current and d) the skewness of the supercurrent flowing through the junction with a high Rashba SOC, $\lambda_{\rm R}=5\hbar v/L$. 
In both Figs.~\ref{fig:NoRashba_1}~c) and d), the dashed black line illustrates the scenario for pristine graphene, where all parameters induced by the substrate are zero. The green solid line indicates the setup with only Rashba SOC $\lambda_{\rm R}=5\hbar v/L$, the blue solid line refers to the case with $\lambda_{\rm R}=5\hbar v/L$ and a smaller VZ SOC $\lambda_{\rm VZ}=\hbar v/L$, and the red solid line represents the condition with $\lambda_{\rm R}=5\hbar v/L$ and a smaller KM SOC $\lambda_{\rm KM}=\hbar v/L$.
Again, the main features of the critical current stem from the proximitized graphene band structure. 
A Rashba SOC term alone modifies the linear band structure of graphene as $E_{\rm G}=\mp \lambda_{\rm R} \pm \hbar v\sqrt{k_x^2+k_y^2 + \lambda_{\rm R}^2}$, introducing a parabolic dispersion close to the charge neutrality point, with the effect of moving and sharpening the Fabry-Perot resonances, as can be seen from the green line in both Figs.~\ref{fig:NoRashba_1}~c) and d).
Including a smaller $\lambda_{\rm VZ}$ (blue solid line) opens a gap of magnitude $2|\lambda_{\rm VZ}|$ in the band structure;  while including a smaller $\lambda_{\rm KM}$ shifts the charge neutrality point of graphene at $-\lambda_{\rm KM}$,  as can be seen from the asymmetry of the critical current as a function of $\mu_0$  (red solid line in Fig.~\ref{fig:NoRashba_1}~c)).
Note that the splitting of the Dirac points due to the Rashba interaction is a second-order effect in the ratio $\lambda_{\rm R}/t$ \cite{Zarea_2009_a}, where $t\sim 2.8$~eV is the graphene hopping parameter. For the values of $\lambda_{\rm R}/t$ that we are considering here, this effect can be safely neglected, even at short junction lengths.

It is noteworthy that the Rashba SOC, despite not significantly impacting the critical current, produces noticeable swings, compared to the intrinsic case, in the skewness of the supercurrent at low chemical potentials. In particular, it can considerably increase the number of transparent modes, with $\mathcal{T}\approx 1$, within the junction.
From Eq.~\eqref{eq:Supercurrent_NoRashba} we see that, in general, a single mode of low transmission, $\mathcal{T} \ll 1$, contributes to the supercurrent as $I_{\mathcal{T}}(\phi) \propto \mathcal{T} \sin(\phi)$ with vanishing skewness, $S \approx 0$, while a transparent mode, $\mathcal{T}\approx 1$, contributes as $I_{\mathcal{T}}(\phi) \propto \sin(\phi/2)$ with maximum skewness, $S\approx 1$.
For a strong Rashba SOC, this leads to highly pronounced peaks in the skewness of the supercurrent, as illustrated in Fig.~\ref{fig:NoRashba_1}~d). 
%
Figure~\ref{fig:MaxSkewRashba}~a) shows $S(\lambda_{\rm R},\mu_0)$ as a function of Rashba SOC and the chemical potential.
Here, the dotted green line represents $\mu_0 = 2\lambda_{\rm R}$; when the chemical potential $\mu_0$ exceeds $2\lambda_{\rm R}$, the upper graphene band is involved in conduction, disrupting the distinct resonances typical of the region with lower chemical potentials.
Moreover, we define the function $\tilde{S}(\lambda_{\rm R}) = \max_{\mu_0}S(\mu_0,\lambda_{\rm R})$, which corresponds to the maximum value of skewness as a function of the chemical potential $\mu_0$ for a fixed value of $\lambda_{\rm R}$. In Figure~\ref{fig:MaxSkewRashba}~a), the red dotted line denotes $\tilde{S}(\lambda_{\rm R})$ as a function of $\lambda_{\rm R}$ in the range $[ 3,7 ] \hbar v/L$. In the main plot of Fig.~\ref{fig:MaxSkewRashba}~b), $\tilde{S}(\lambda_{\rm R})$ is reproduced within the range $\lambda_{\rm R}\in [3,7] \hbar v/L$, and there is a global maximum around $\lambda_{\rm R \blacktriangle}\approx4.7 \hbar v/L$.
The inset of Fig.~\ref{fig:MaxSkewRashba}~b) shows the nearly step-like transmission probabilities as a function of the momentum $k$ at the value of the chemical potential $\mu_0$ which maximizes the skewness for $\lambda_{\rm R}=\lambda_{\rm R \blacktriangle}$.
%
Thus, by influencing $\lambda_{\rm R}$, such as through the application of a transverse electric field \cite{Min_2006_a,Gmitra_2009_a,Gmitra_2015_a,Omar_2018_a}, one can potentially enhance the higher harmonics into the CPR of a wide and ballistic GJJ by increasing its skewness.

\section{Conclusions}\label{sec:conclusions}

In this work, we studied the current-phase relation in a ballistic GJJ in which the inner graphene layer is subjected to different spin-orbit coupling interactions because of the proximity effect with a substrate. We focused on the short and wide limits, at zero temperature. We found an explicit analytical expression for the combined effects of the onsite scalar potential, Kane-Mele, and valley-Zeeman spin-orbit couplings. This expression for the transmission probabilities can be written in the same form as the intrinsic case with a suitable renormalization of the two parameters $\alpha_{\rm s \tau}$ and $\gamma_{\rm s \tau}$. In particular, the valley-Zeeman term combines with the chemical potential to produce an effective spin-valley Fermi level. Instead, the Kane-Mele interaction combines with the onsite scalar potential to make a spin-valley-dependent gap, which lowers the supercurrent transport for small chemical potential values.

The effects of a Rashba spin-orbit coupling were studied numerically for experimentally relevant cases that include a high Rashba interaction with a smaller secondary one. Again, most of the results about the supercurrent phase relation stem from band structure modifications of the proximitized graphene; a small valley-Zeeman term opens a gap, while a small Kane-Mele term shifts the charge-neutrality point of graphene. Interestingly, we found that the Rashba interaction can sensibly boost the number of transparent modes through the junction, producing heavy swings in the skewness of the current-phase relation, which could have applications for devices needing tunable harmonic content.

\acknowledgements
The authors thank G.G.N. Angilella and V. Varrica for fruitful comments on various stages of this work. E.P. and F.M.D.P. thank the PNRR MUR project PE0000023-NQSTI.
E.P. acknowledges support from COST Action CA21144 superqumap. F.B. acknowledges support from the project PRIN 2022 - 2022XK5CPX (PE3) SoS-QuBa - "Solid State Quantum Batteries: Characterization and Optimization". 
G.F. thanks for the support ICSC - Centro Nazionale di Ricerca in High-Performance Computing, Big Data and Quantum Computing under project E63C22001000006 and Universit\`a degli Studi di Catania, Piano di Incentivi per la Ricerca di Ateneo, project TCMQI.

\bibliographystyle{mprsty}
\bibliography{Bibliography}

\end{document}